\documentclass[letterpaper, 10 pt, conference]{ieeeconf}  

\IEEEoverridecommandlockouts                              
\usepackage{graphics} 
\usepackage{epsfig} 
\usepackage{mathptmx} 
\usepackage{times} 
\usepackage{amsmath} 
\usepackage{amssymb}  
\usepackage{bm,subfigure,color}
\newenvironment{myindentpar}[1]%
 {\begin{list}{}%
         {\setlength{\leftmargin}{#1}}%
         \item[]%
 }
 {\end{list}}

\title{\LARGE \bf
Safe Protocol for Controlling Power Consumption by a Heterogeneous Population of Loads
}

\author{Soumya Kundu and Nikolai Sinitsyn
\thanks{S. Kundu is with the Department of Electrical Engineering and Computer Science,
        University of Michigan, Ann Arbor, MI 48109, USA
        {\tt\small soumyak@umich.com}}%
\thanks{N. Sinitsyn is with the Theoretical Division, Los Alamos National Laboratory,
        Los Alamos, NM 87544, USA
        {\tt\small sinitsyn@lanl.gov}}%
}

\begin{document}

\maketitle
\thispagestyle{empty}
\pagestyle{empty}

\begin{abstract}
Recent studies  on control of aggregate power of an ensemble of thermostatically-controlled-loads (TCLs) have been concentrated on shifting the temperature set points of each TCL in the population. A sudden shift in the set point, however, is known to be associated with undesirable power oscillations which require closed-loop control strategies to regulate the aggregate power consumption of the population. In this article, we propose a new approach which we term as a ``safe protocol'' to implement the shift in temperature set point. It is shown analytically and verified numerically that by shifting the set point ``safely'' the aggregate power consumption can be changed to a different value within a time frame of the order of a TCL's cycle duration and avoid the undesired oscillations seen otherwise in a ``sudden'' shift. We  discuss how the excess aggregate energy transferred under a safe shift in the set point could potentially mitigate the burden due to abnormal energy generation within a short time span.


\end{abstract}

\section{INTRODUCTION}

With increasing level of penetration of renewable energy sources into the power grid, focus of many recent studies has been on the potential of electrical loads in reducing generation-consumption mismatch. Higher {\it intermittency} and {\it non-dispatchability} associated with increased dependence on renewable energy sources can be better taken care of by electrical loads than by conventional generators (\cite{strbac,klobasa}) which have much higher response time and usually offer more expensive and environmentally damaging way of mitigating fluctuations in renewable generation. Thermostatically-controlled-loads (TCLs) account for about $50\%$ of electricity consumption in the United States \cite{EIA} and have been studied for their capability to perform generation-balancing ancillary services \cite{callaway,hiskens}. 

For simplicity, we assume that our TCLs are air conditioners in houses of a large city. Such TCLs are working in cycles by switching between ``ON'' (drawing electrical power) and ``OFF'' (not drawing any power) states. We would assume that the power distribution authority has the ability to interfere in the process of operation of TCLs, which it intends to use for offsetting power fluctuations in the grid. Traditional focus on applied thermal load control has been  either on direct interruption of power or on a sudden shift of the {\it hysteresis-deadband} around setpoint temperature of the TCLs \cite{ihara}-\cite{ucak}. The former approach can affect customer interest if such disruptions happen for several hours in a hot day.  Hence, the approach that is based on the centralized hysteresis-based control of the TCLs' deadband positions appears more attractive \cite{callaway,pscc}. In this case, power fluctuations in the grid are compensated by subtle ($\sim 0.1^oC$) changes in the thermostat setpoint temperature. It is expected that such minuscale variations in thermostat setpoint will remain almost unnoticed by the customers. This minimally invasive approach should attract more customers and can be considered preferable for a large-scale commercial implementation. The idea of shifting the operation band of TCLs has faced, however, with intrinsic problem of synchronization of TCLs. Simulations of a behavior of an ensemble of TCLs show that instead of eliminating unwanted power fluctuations in the grid, uniform ``sudden'' shift of setpoints of all TCLs can, without closed-loop feedback control, synchronize the TCLs and lead to strong unwanted power oscillations in the grid \cite{callaway,pscc}. 

In this article, we present a method to implement the shift in temperature setpoint of the TCLs, which we will refer to as the ``safe protocol'', that completely eliminates the TCL synchronization problem. We introduce a pair of ``transition points'' that govern the thermal dynamics of TCLs during the shift of setpoint temperature. 
Section~\ref{problem_statement} describes the thermal dynamics of a large population of TCLs and the oscillation associated with setpoint change. Section~\ref{safe_sec} details the working of ``safe protocol'' while section~\ref{power_profile} computes an aggregated power and energy consumption during this process. Simulation results are presented in Section~\ref{numerical} while Section~\ref{conclusion} concludes with a brief summary of this work and future research direction.

\section{PROBLEM DESCRIPTION}\label{problem_statement}
\begin{figure}[thpb]
\centering
\includegraphics[width=2.5in]{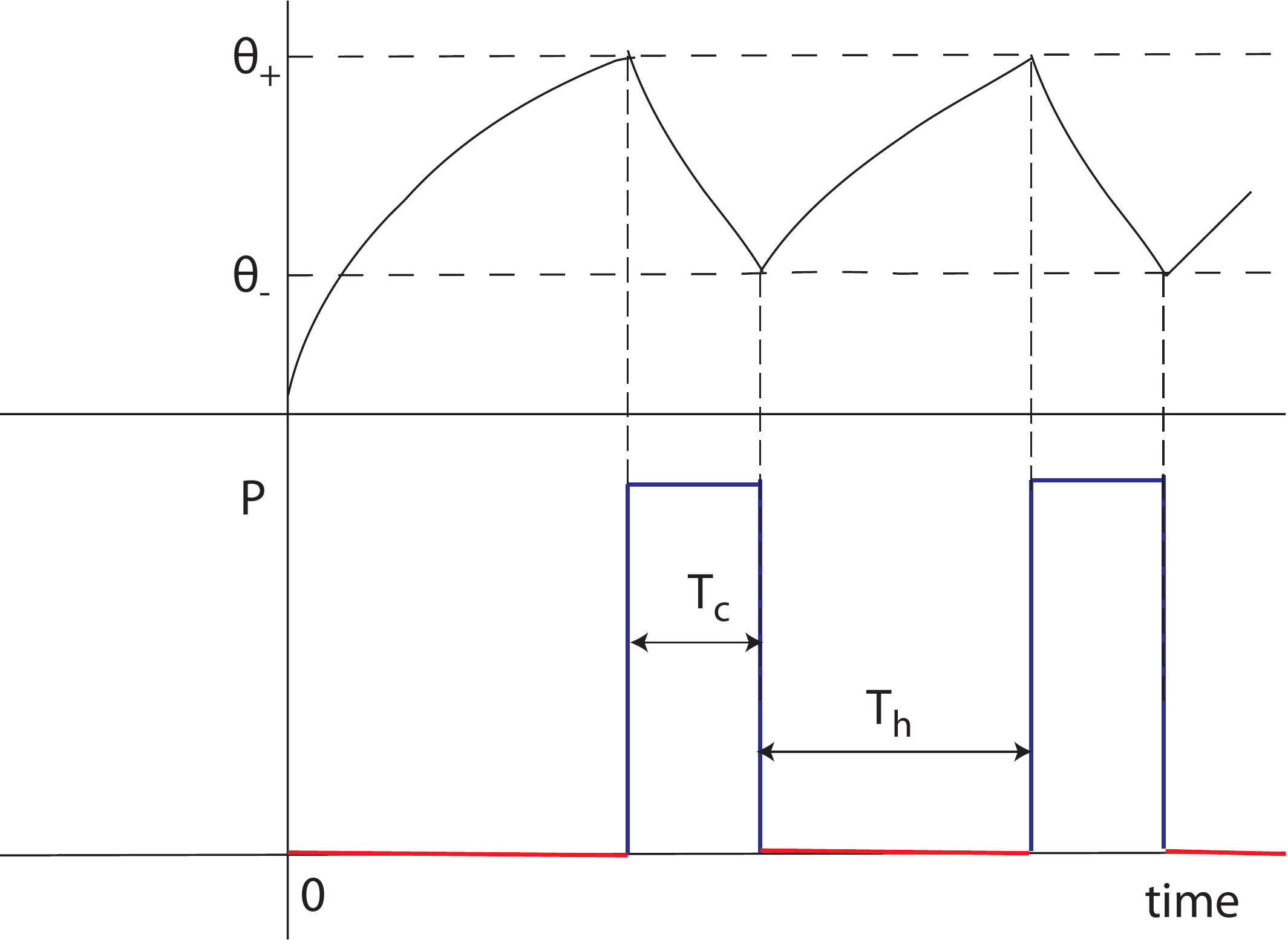}
\caption{Dynamics of temperature of a thermostatic load.} \label{dynamics}
\end{figure}
Consider a ``minimal model'' of a single cooling type TCL \cite{pscc}. The temperature, $\theta$, of a TCL drops when it draws power (in the ON state) and increases when the TCL is not drawing any power (in the OFF state): 
\begin{equation}
\dot{\theta} = \left\lbrace \begin{array}{ll} -\frac{1}{CR}\left(
  \theta - \theta_{amb} +PR \right), & \text{ON state} \\ &
  \\ -\frac{1}{CR}\left( \theta - \theta_{amb} \right), &
  \text{OFF state}	\end{array} \right. \label{micro}
\end{equation}
where $\theta_{amb}$ is the ambient temperature, $C$ is the thermal
capacitance, $R$ is the thermal resistance, and $P$ is the power drawn
by the TCL when in the ON state. 
The dynamics forces a TCL's temperature into a hysteresis deadband from $\theta_-$ to $\theta_+$ around the setpoint temperature $\theta_s=\left(\theta_-+\theta_+\right)/2$. A TCL switches its state from OFF to ON when its temperature increases to $\theta_+$ and from OFF to ON when temperature drops to $\theta_-$.
Solution of Eqs. (\ref{micro}) with such boundary conditions is shown in Fig.~\ref{dynamics}~(top). Fig.~\ref{dynamics}~(bottom) shows that, at steady state, power consumption by a TCL switches between a constant and zero values, respectively in ON and OFF states. 

\begin{figure*}[thpb]
\centering
\subfigure[]{
\includegraphics[width=2in]{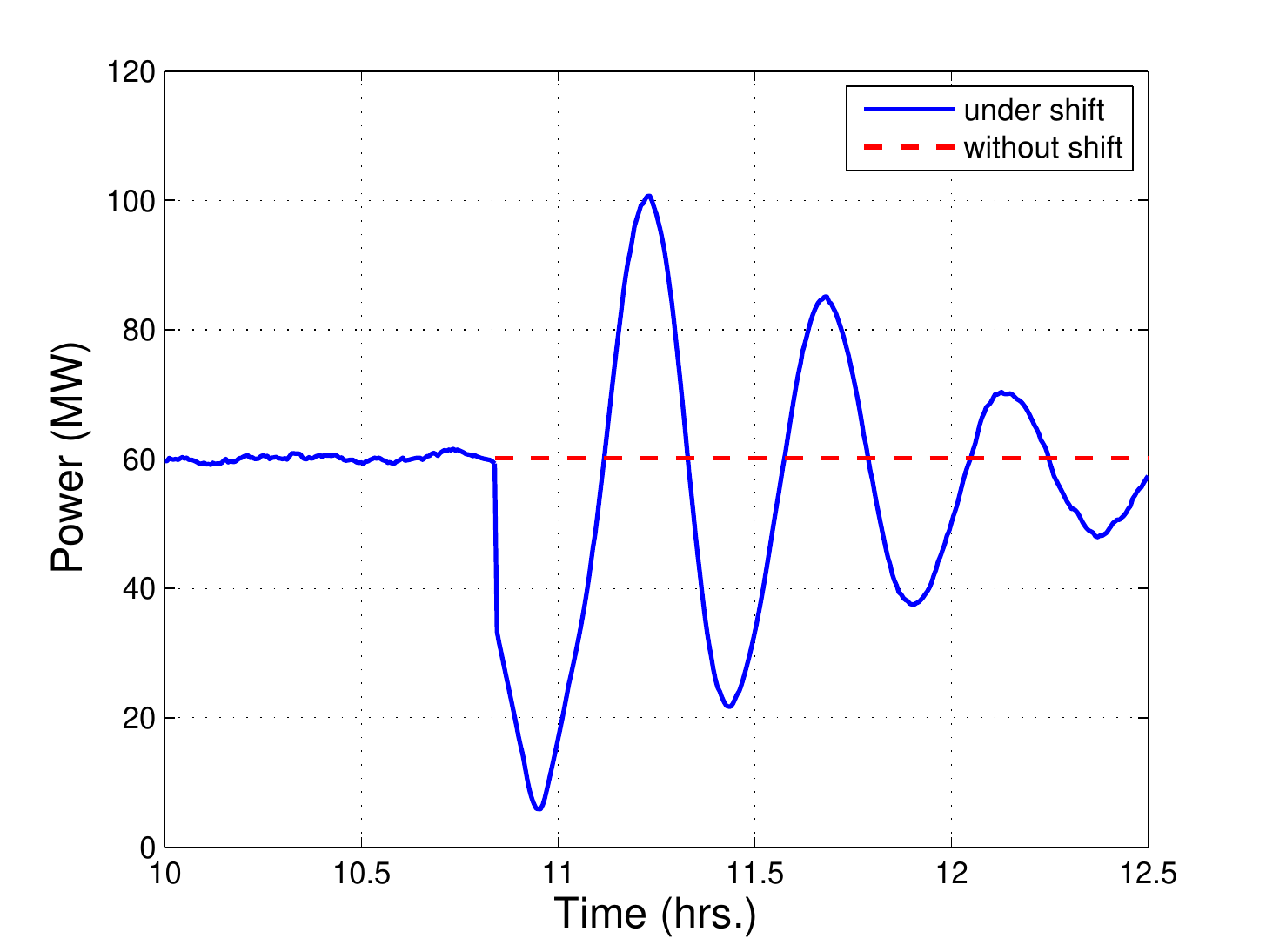}\label{normal_R}
}\quad
\subfigure[]{
\includegraphics[width=2in]{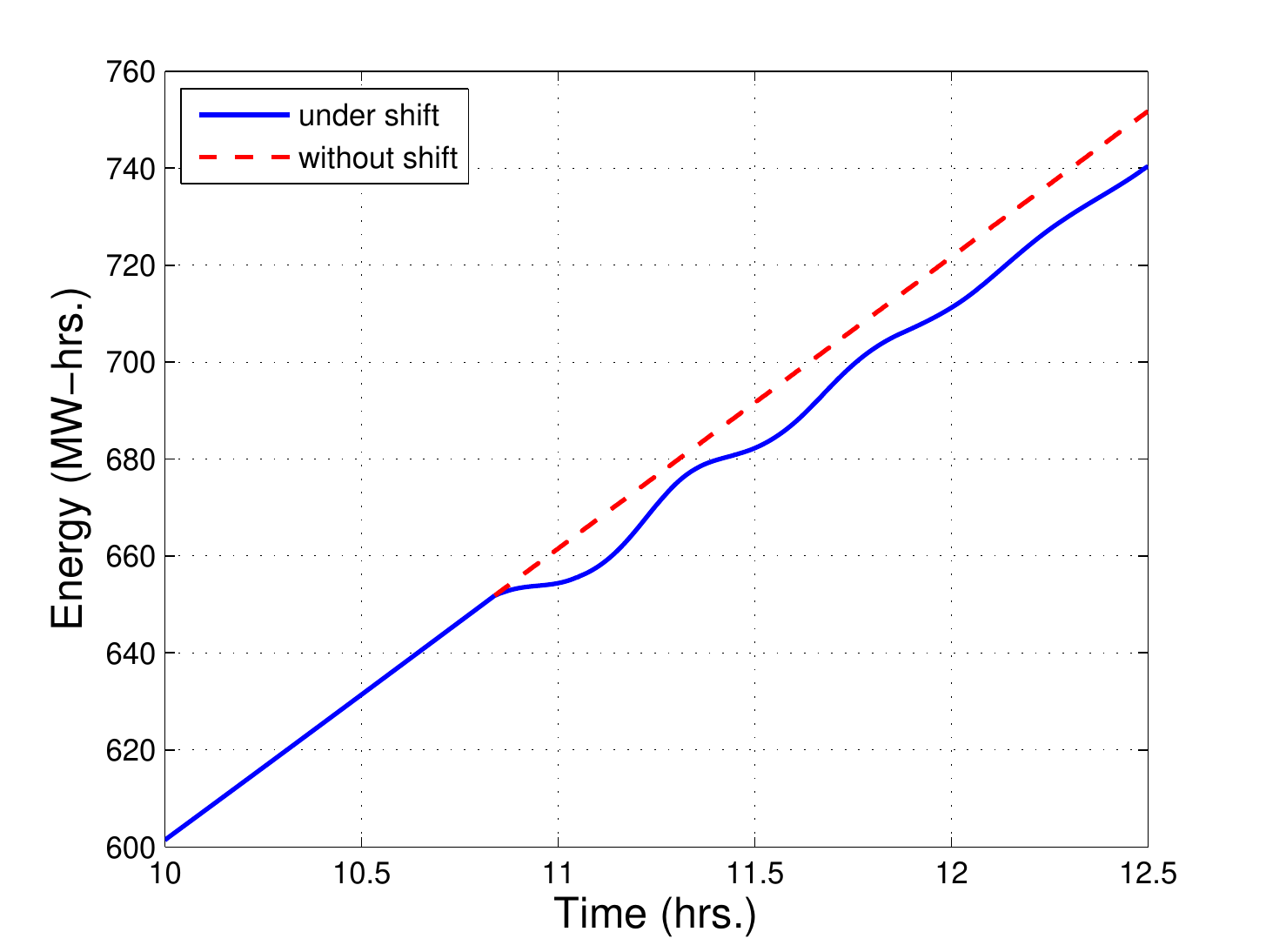}\label{normal_R_E}
}\quad
\subfigure[]{
\includegraphics[width=2in]{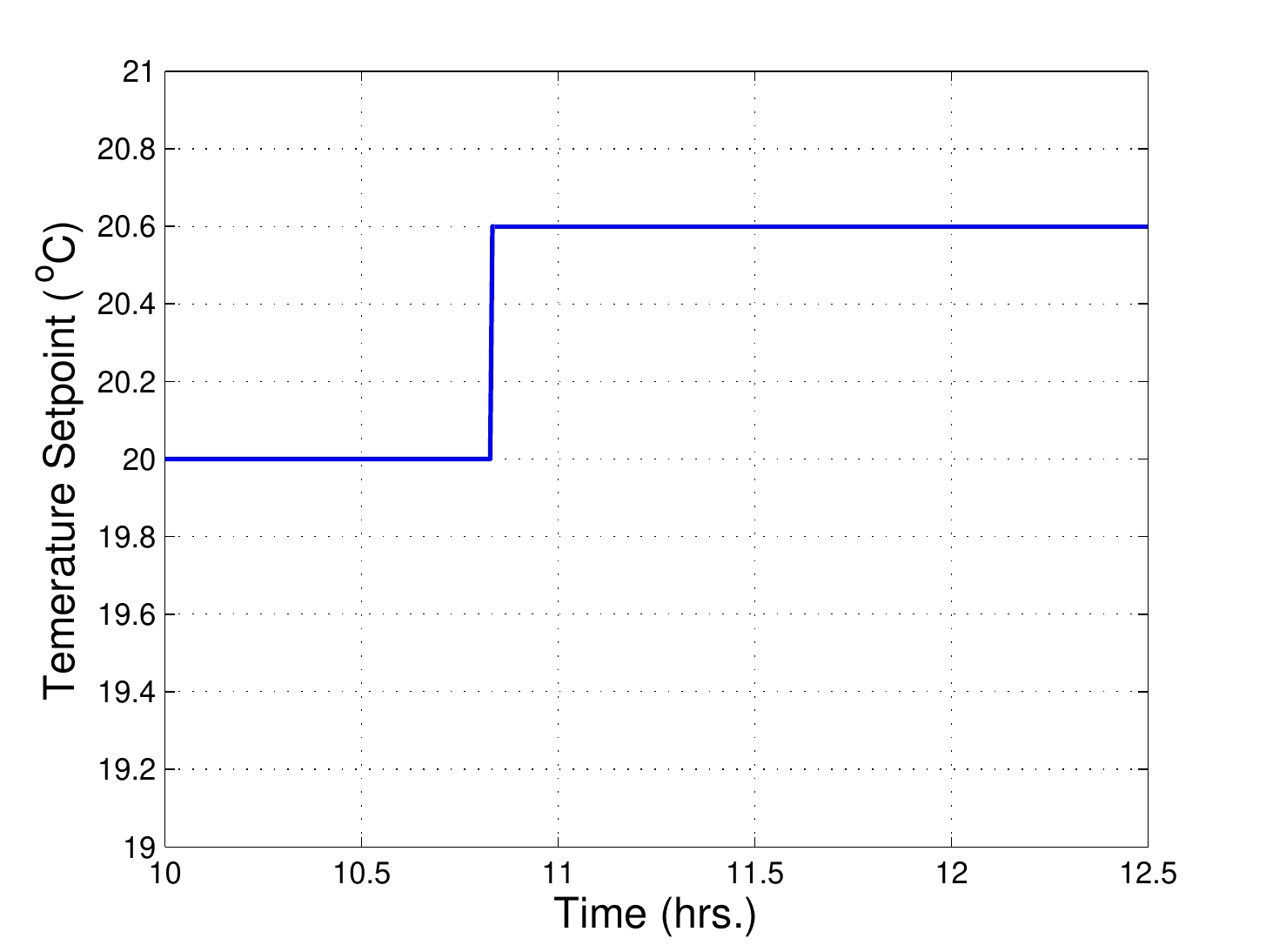}\label{input_R}
}\caption[Optional caption for list of figures]{Power and energy consumption profiles under a ``sudden'' shift to higher setpoint temperature.}
\label{normal_increase}
\end{figure*}
\begin{figure*}[thpb]
\centering
\subfigure[]{
\includegraphics[width=2in]{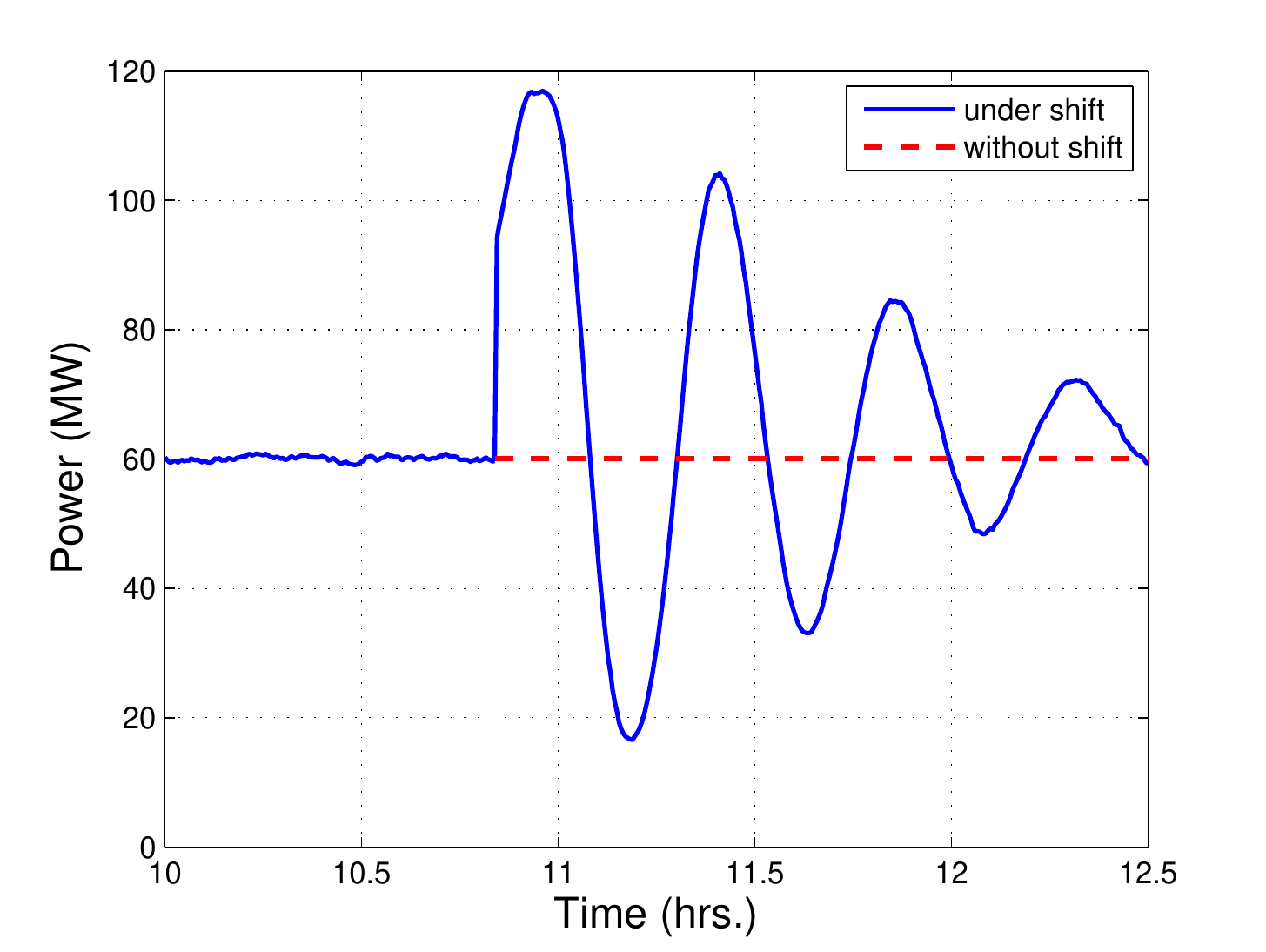}\label{normal_L}
}\quad
\subfigure[]{
\includegraphics[width=2in]{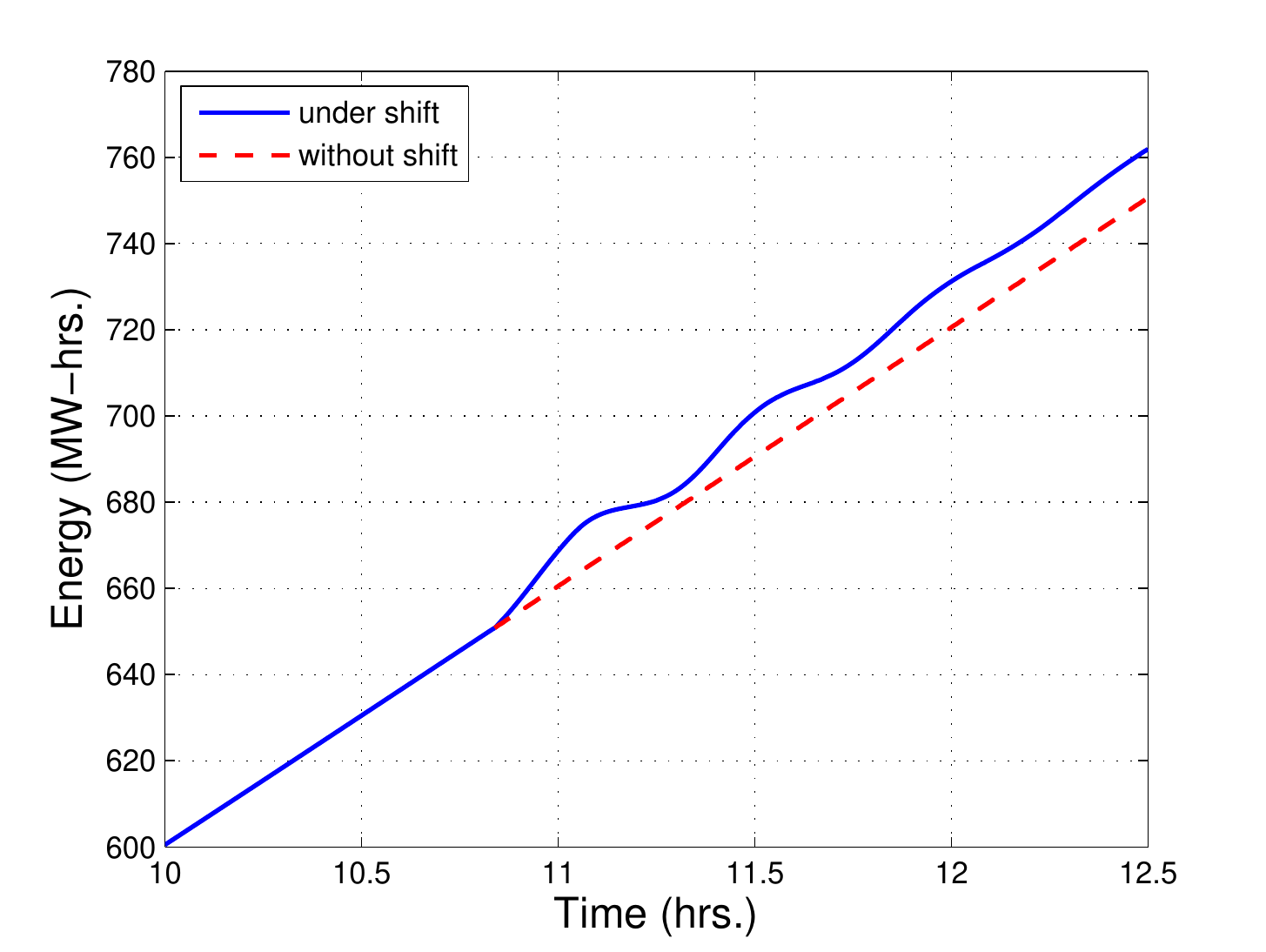}\label{normal_L_E}
}\quad
\subfigure[]{
\includegraphics[width=2in]{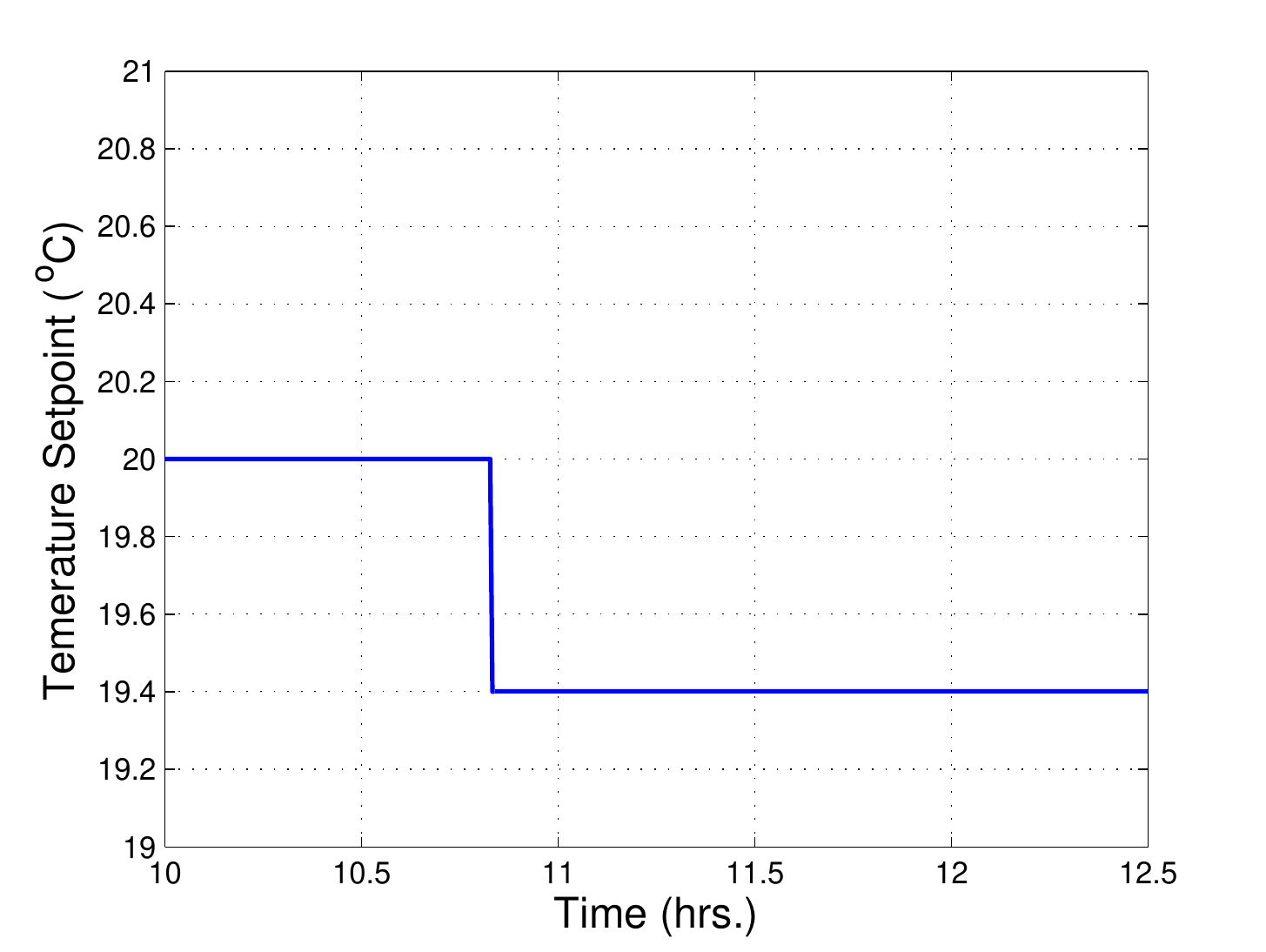}\label{input_L}
}\caption[Optional caption for list of figures]{Power and energy consumption profiles under a ``sudden'' shift to lower setpoint temperature.}
\label{normal_decrease}
\end{figure*}

For a large {\it heterogeneous} ($C,R$ and $P$ values are different across the population) ensemble of TCLs, the aggregate power demand attains a steady state value \cite{callaway}. In steady state, temperatures of all the TCLs would lie within the hysteresis deadband. Denoting $T_c$ and $T_h$ as the mean times spent in ON and OFF states, respectively, the steady state probability density function of the ON state TCLs, $f_1\left(\theta\right)$, and of the OFF state TCLs, $f_0\left(\theta\right)$, can be estimated as, \cite{pscc}:
\begin{eqnarray}
\label{f1f0_estimate}
f_1(\theta) &=& \frac{CR}{(T_c+T_h)(PR+\theta-\theta_{amb})}, \quad \forall\theta\in[\theta_-,\theta_+] \nonumber \\
\& \quad f_0(\theta) &=& \frac{CR}{(T_c+T_h)(\theta_{amb}-\theta)}, \quad \forall\theta\in[\theta_-,\theta_+] 
\end{eqnarray}
where,
\begin{eqnarray}\label{TcTh}
T_c &=& {CR} \ln \left( \frac{PR+\theta_{+}-\theta_{amb}}
{PR+\theta_{-}-\theta_{amb}} \right) \nonumber\\
T_h &=& {CR} \ln \left( \frac{\theta_{amb} - \theta_{-}} {\theta_{amb} -
  \theta_{+}} \right) \nonumber\\
  \& \quad T_{tot} &=& T_c + T_h
\end{eqnarray}
\begin{figure}[thpb]
\begin{center}
\includegraphics[width=3in]{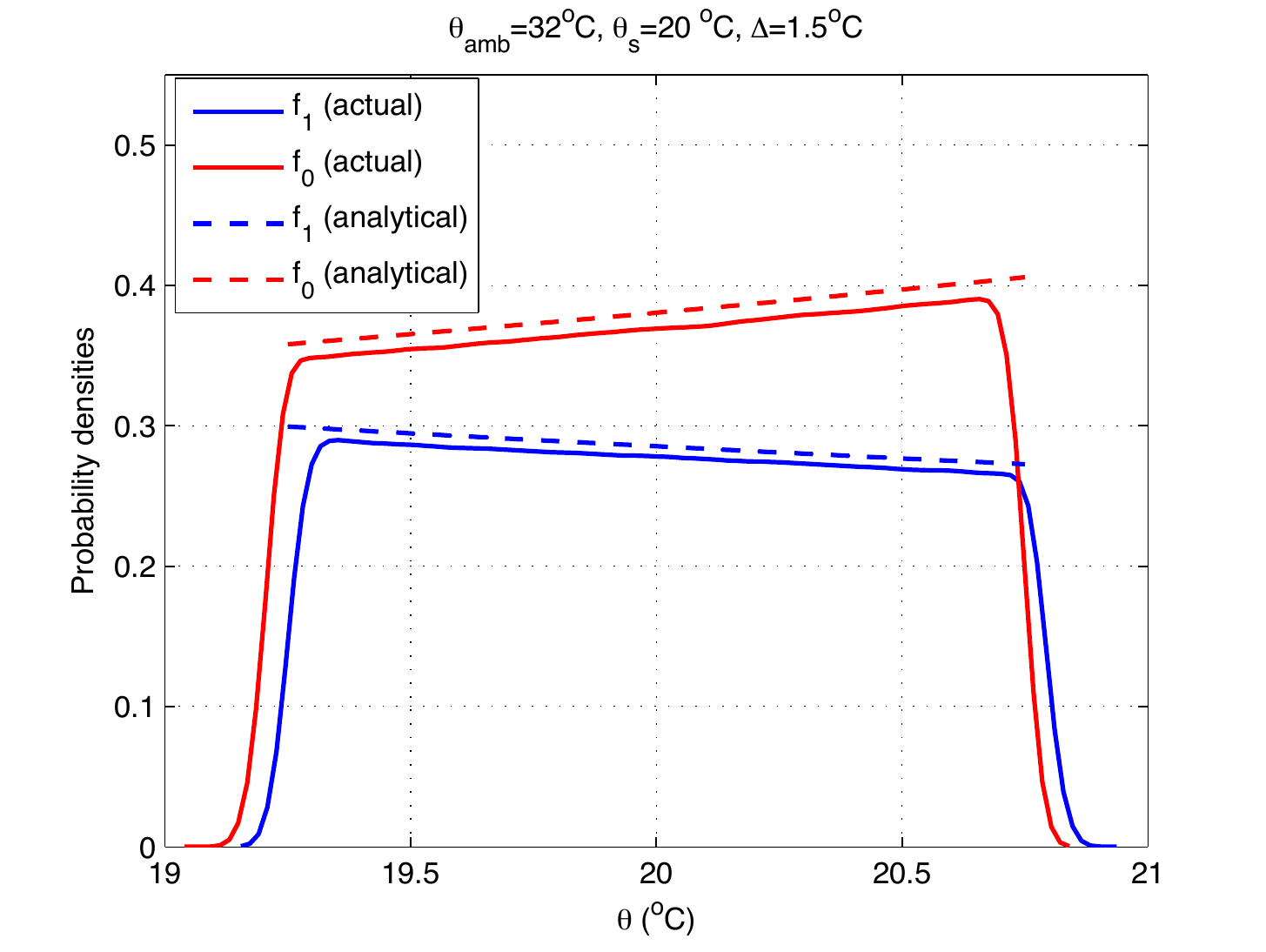}
\caption{Steady state probability densities. Small but nonzero values of probabilities beyond the deadband position are due to temperature fluctuations. } \label{f1f0}
\end{center}
\end{figure}
Fig.~\ref{f1f0} shows the steady state probability distribution for a heterogeneous ensemble of TCLs\footnote{$N=10,000, \theta_{amb} = 32~^oC$; $C,R$ and $P$ follow lognormal distributions with mean $1~kWh/^oC, 2 ~^oC/kW$ and $14~kW$, respectively and a standard deviation of $\sigma_p=0.07$ of the corresponding mean; a zero-mean Gaussian noise with standard deviation of $0.052^oC/min^{0.5}$ was added to (\ref{micro}).}, where the setpoint temperature is $20~^oC$ with deadband width $\Delta = \left(\theta^+-\theta^-\right)=1.5~^oC$. Fig.~\ref{normal_increase}-\ref{normal_decrease} show typical responses to a step change in temperature setpoint. Before the ``sudden'' change of the position of the deadband, the power consumption is almost uniform with small fluctuations resulting from a finite size of the TCL population and noise. When the temperature setpoint is suddenly shifted to a different value (keeping the width $\Delta$ constant), at around $10.8~hrs.$, the aggregate power demand breaks into a damped oscillatory mode before settling down onto a different steady state value. Figs.~\ref{normal_R} and~\ref{normal_R_E} show the aggregate power and energy consumption corresponding to the setpoint increase in Fig.~\ref{input_R}, while Figs.~\ref{normal_L} and~\ref{normal_L_E} correspond to setpoint decrease in Fig.~\ref{input_L}. The power profiles show large oscillations which are also reflected in the energy consumption profiles. 

While, for a fairly {\it homogeneous} system, a centralized control on a physically based model can be used to force the aggregate power into tracking certain reference signals \cite{pscc}, these parasitic oscillations in a {\it heterogeneous} population are hard to remove by centralized control signals because of randomness of parameters in the ensemble. Quite often, changes of the temperature setpoint in a range of $0.5~^oC$-$1~^oC$ would be desirable from utility perspective. At such amplitudes, however, the parasitic disturbances of the system become a considerable problem that cannot be addressed by a linear control approach, which performs well for disturbances that are smaller than  $0.1~^oC$. It is thus desirable to design a ``shift'' mechanism that eliminates the problem of synchronization and related power oscillations and yet exercises certain open-loop control over the energy consumption in a short duration (less than a time period of oscillation).

\section{THE SAFE PROTOCOL}
\label{safe_sec}
In this section we explain how a ``safe protocol'' would work. In a ``sudden shift'' of temperature setpoint, the new set of deadband limits are instantly applied to all the TCLs in the population which leads to a sudden jump in the aggregate power consumption. To avoid this, the shift of deadband limits will be applied to all the TCLs gradually in a certain way such that the new set of deadband limits are operative within less than a time period. Before we start to explain the mechanism, we need to clarify a concept of ``transition points'' to be used here:

\begin{myindentpar}{0.3cm}
``Transition points'' are intermediate hysteresis deadband limits that are operative starting from the instant the shift is initiated until the new set of deadband limits are applied across the population. Let $\theta_-^0$ and $\theta_+^0$ be the lower and upper deadband limits before shift. If the temperature setpoint is to be increased by an amount $\delta$, then the transition points would be $\theta_-^0$ and $\left(\theta_+^0+\delta\right)$. While for a decrease in setpoint by an amount $\delta$, the transition points would be $\left(\theta_-^0 - \delta\right)$ and $\theta_+^0$.
\end{myindentpar}
 
 Let a signal be received by a TCL to shift the temperature setpoint by $\delta$ ($\delta>0$ for increase, while $\delta<0$ for decrease). At this moment this TCL can be anywhere on the original cycle. The ``safe protocol'' is constructed this way:
 \begin{enumerate}
 \item This TCL continues to stay in its present state (ON or OFF) until it hits one of the transition points. 
 \item Once it has reached one of the transition points, the new pair deadband limits, $\theta_-=\left(\theta_-^0+\delta\right)$ and $\theta_+=\left(\theta_+^0+\delta\right)$, start to govern its thermal dynamics. 
 \end{enumerate}
 
 \begin{figure*}[thpb]
\centering
\subfigure[Initiation of the process, $t=0$]{
\includegraphics[width=2in]{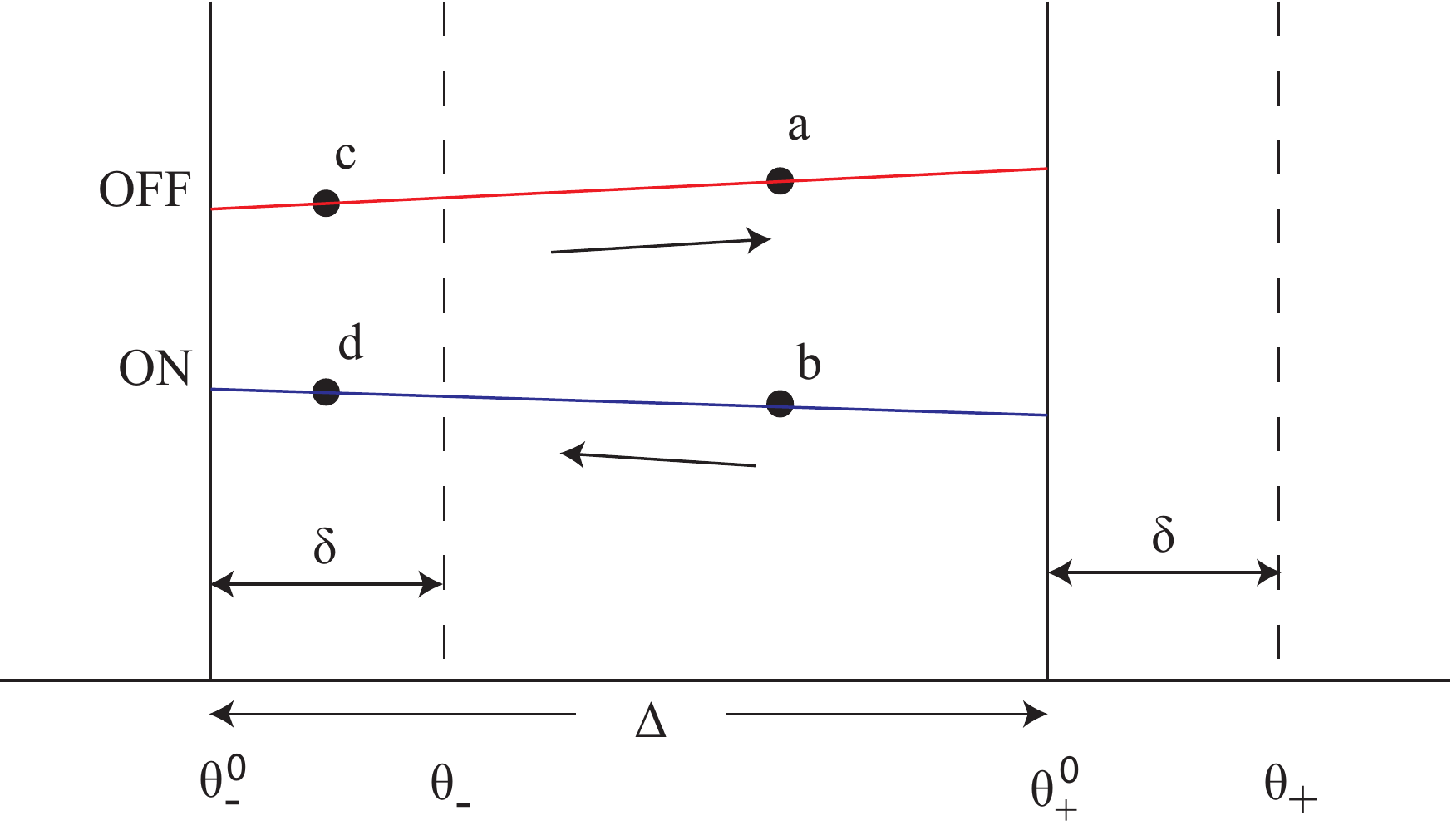}\label{R1}
}\quad
\subfigure[$0<t<\tau_1$]{
\includegraphics[width=2in]{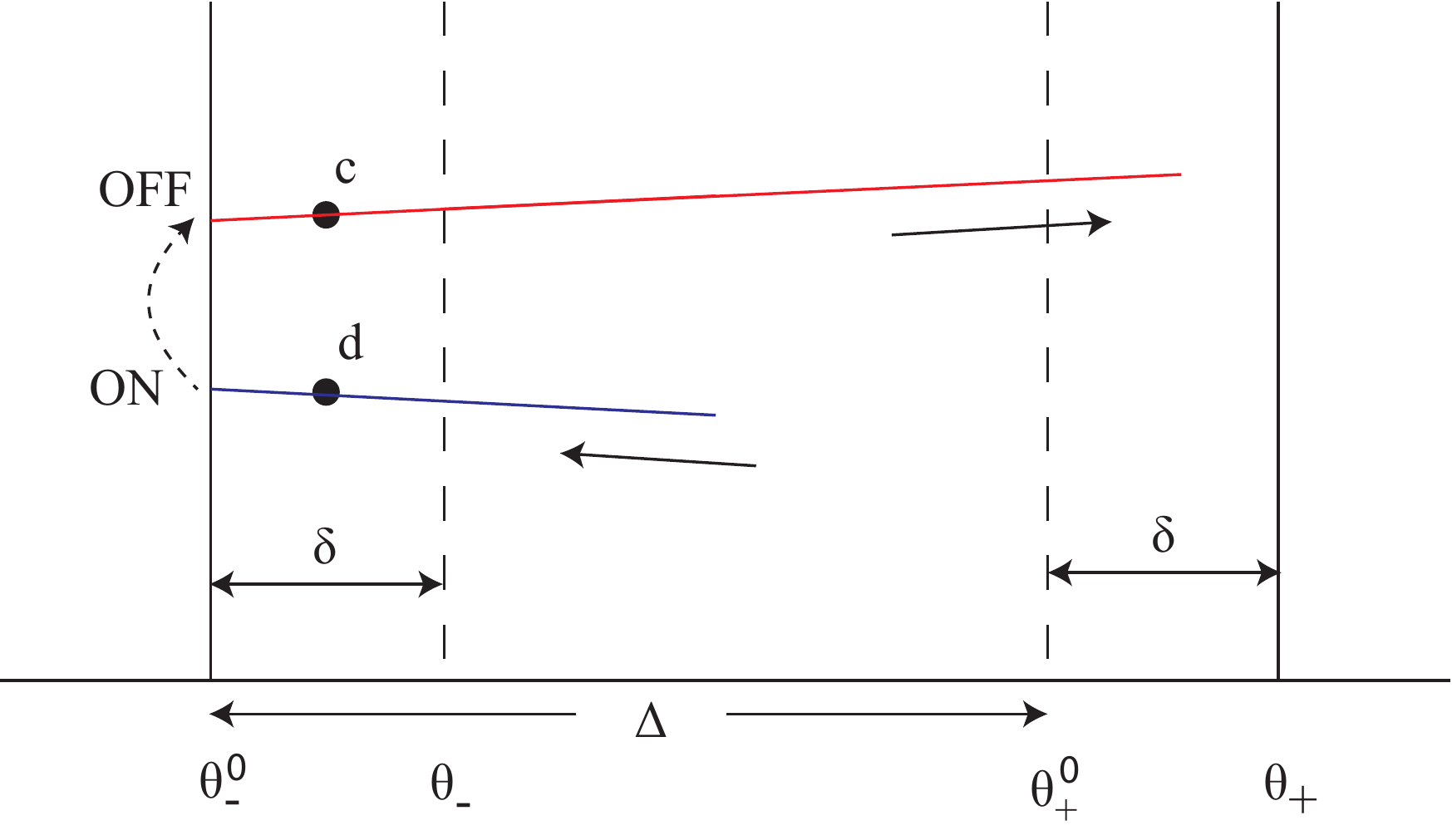}\label{R2}
}\quad
\subfigure[$\tau_1<t<\tau_2$]{
\includegraphics[width=2in]{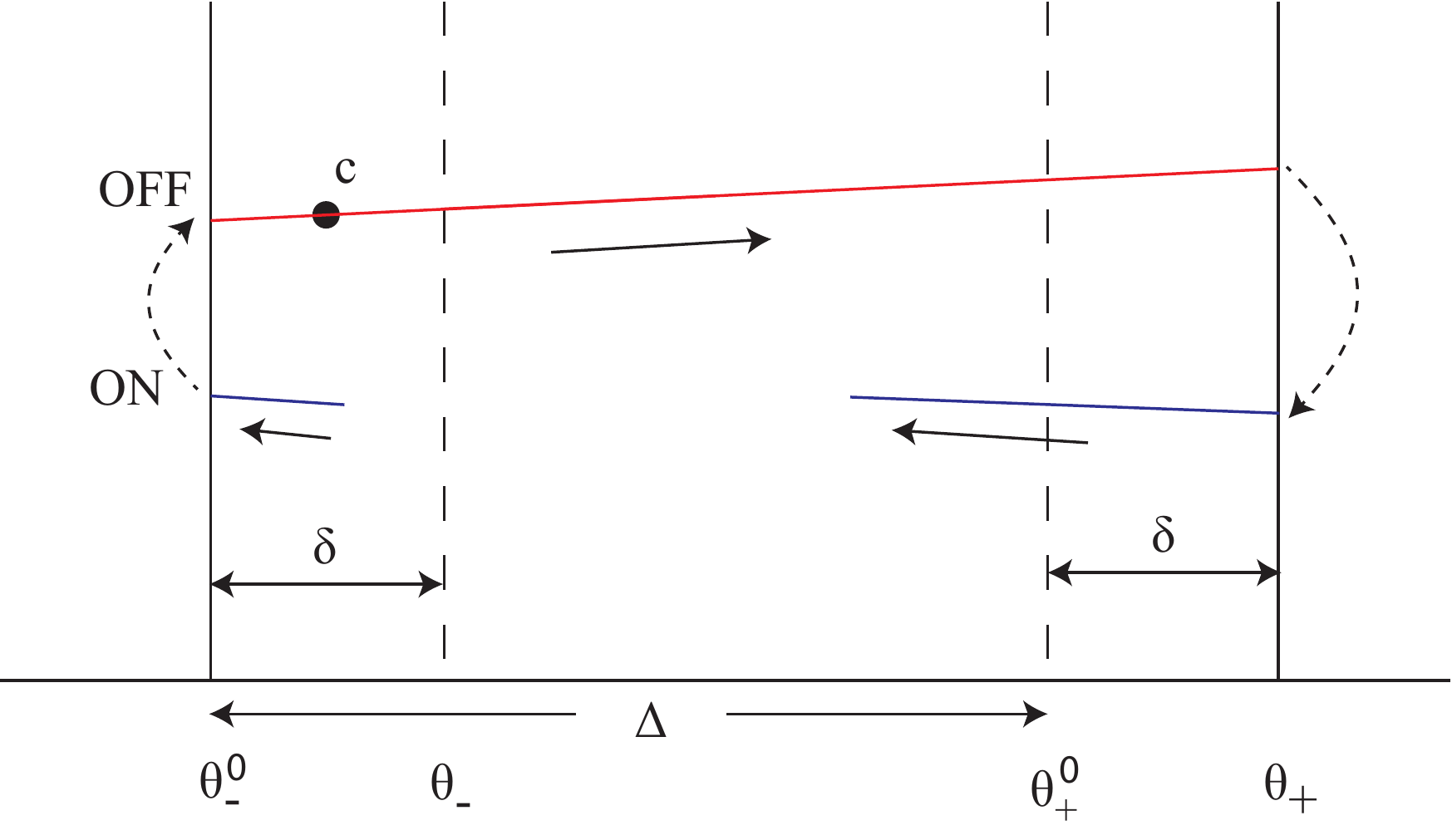}\label{R3}
}\quad
\subfigure[$\tau_2<t<\tau_3$]{
\includegraphics[width=2in]{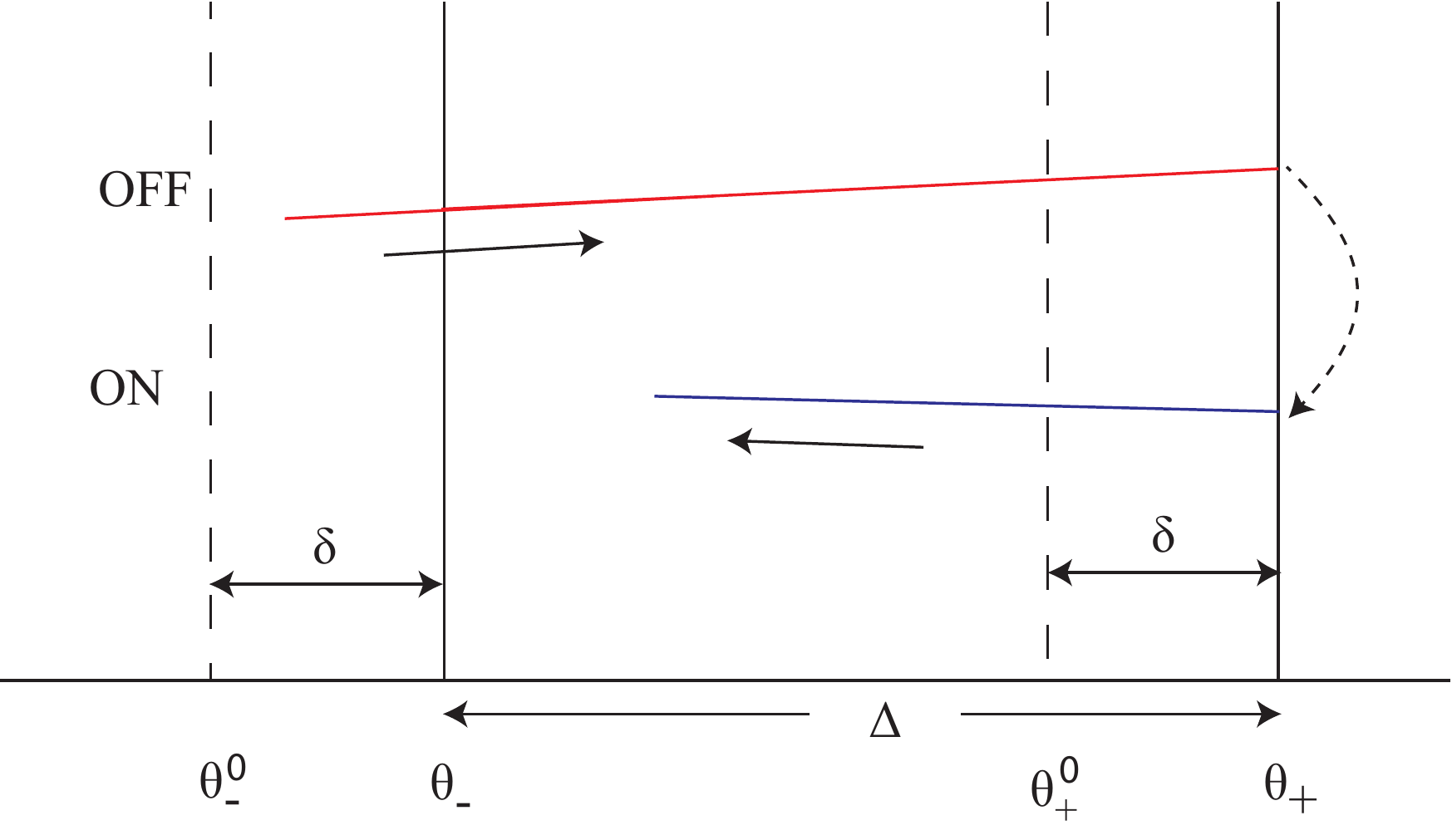}\label{R4}
}\quad
\subfigure[Completion of the process, $t=\tau_3^+$]{
\includegraphics[width=2in]{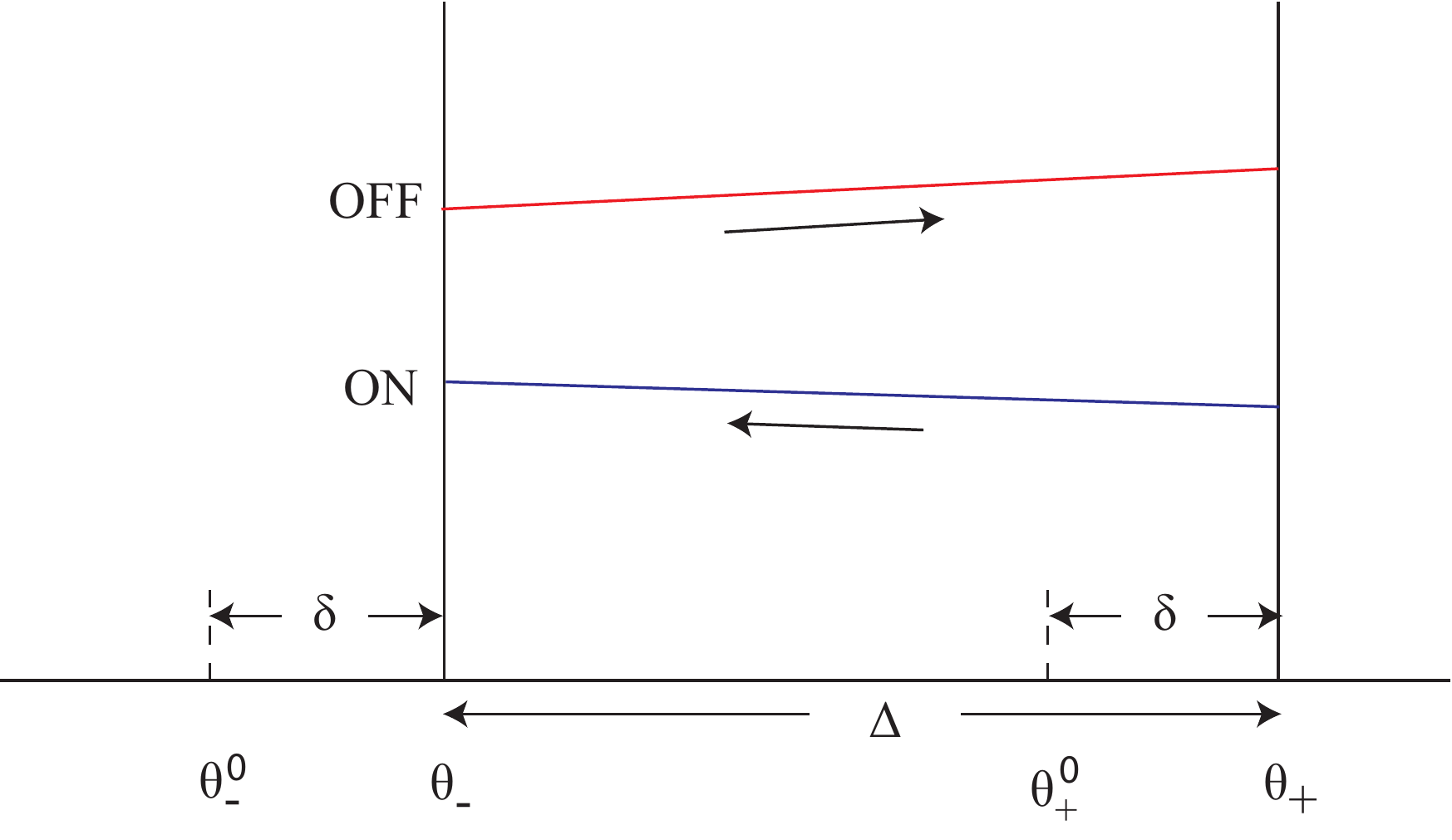}\label{R5}
}
\caption[Optional caption for list of figures]{Changes taking place in the probability density of ON and OFF profiles when $\delta>0$: (a) Initial equilibrium distribution. (b) No switching at $\theta_+^0$ while switching continues at $\theta_-^0$. (c) Switching starts at $\theta_+$ while switching continues at $\theta_-^0$ until {\it all} of the initially ON TCLs have switched to ON once. (d) Both new deadband limits are now activated and the first set of ON TCLs are approaching $\theta_-$. (e) New equilibrium distribution is attained within the new deadband limits.
}
\label{safe_protocol}
\end{figure*}
Fig~\ref{safe_protocol} illustrates distinct stages of evolution of the probability distribution over temperature in an ensemble of TCLs after it is perturbed by the ``safe shift'' of setpoint (in this case, an increase in setpoint). Before the initiation of the process, all the TCLs had their temperature lying between $\theta_-^0$ and $\theta_+^0$ (Fig.~\ref{R1}). Once the initiated each TCL continues with its state (ON or OFF) till its temperature hits either of the transition points, $\theta_-^0$ or $\theta_+^0+\delta$, as in Figs.~\ref{R2}-\ref{R3}. Beyond time $t=\tau_2$, the new set of deadband limits, $\theta_-=\left(\theta_-^0+\delta\right)$ and $\theta_+=\left(\theta_+^0+\delta\right)$, are applied to {\it all} the TCLs in the population (Fig.~\ref{R4}). Finally the population settles itself to a distribution within the new deadband limits (Fig.~\ref{R5}). 

If the period of the new cycle coincides with the period of the original cycle, $T_{tot}$, then after time $T_{tot}$ the ensemble of TCLs that  is uncorrelated before the control signal will be in the uncorrelated state. This follows from the fact that the rate of TCLs arrivals at the transition point is constant  and all TCLs pass through it during $T_{tot}$ - just as in the case of uncorrelated steady state with the new deadband position. Hence, the final distribution of TCLs over temperature is the same as it would be at steady state conditions with the new position of the deadband. This ``strict version'' of safe protocol requires to install some sort of a ``smart meter'' on the TCL, which would learn the local parameters $C,R,\theta_{amb},P$ and original deadband width $\Delta_0$ and accordingly set the new deadband width $\Delta$ so that the new time period $T_{tot}$ equals original time period $T_{tot}^0$. 

However it is observed that the relative change in time period, $\eta_T$, is generally very small. Using the parameters in Section~\ref{problem_statement}, the change in time period for a shift $\delta = 1^oC$ comes out to be
\begin{eqnarray}\label{eta_T}
\eta^{T} &=& \frac{T_{tot}-T_{tot}^0}{T_{tot}^0} \approx 0.015 
\end{eqnarray}
In our example, with initial setpoint temperature at $20^oC$ and ambient temperature at $27-35^oC$, a shift of the order of $1^oC$ would cause only about $2\%$ change in the time period. Considering local temperature fluctuations and variation of parameters with time, the impact of such change of time period ($\eta_T\ll 1$) on successful implementation of safe protocol would be imperceptible. $\eta_T$ would be significant when setpoint temperature is either very close to or quite below ambient temperature, neither situation is not controllable as almost all the TCLs are either OFF or ON. Hence we propose a ``light version'' of safe protocol would be sufficient, where local deaband adjustments to maintain a strictly same time period is not necessary. Fig.~\ref{safe_protocol} depicts this scenario when the only setpoint shifts but the deadband width remains fixed allowing the time period to change.

\section{POWER AND ENERGY CONSUMPTION}\label{power_profile}

In this section, a time response of the aggregate power, $P_{tot}(t)$, and energy consumption by the TCLs to a shift in the temperature setpoint under the safe protocol is derived assuming homogeneity of parameters. 


In the steady state, the {\it out-flows} and {\it in-flows} of both the ON state probability and the OFF state probability remain equal. 
Assuming $t$ be the time a TCL (in a homogeneous population) takes to heat up from $\theta_t$, $\theta_t\in(\theta_-,\theta_+)$ to $\theta_+$, the number of TCLs going from OFF to ON state in time $t$ would be
\begin{eqnarray}\label{flow_01}
\Delta N_{0,1}^t &=& N\int_{\theta_t} ^{\theta_+} f_0(\theta)d\theta, \quad \theta_t = \theta_{amb}-\left(\theta_{amb}-\theta_+\right)e^{\frac{t}{CR}}\nonumber\\
 &=& N\frac{t}{T_c+T_h} = N\frac{t}{T_{tot}} \quad (using\;(\ref{f1f0_estimate}))
\end{eqnarray}
By similar calculation for the ON state, the number of TCLs going from ON to OFF in time $t$ is
\begin{eqnarray}\label{flow_10}
\Delta N_{1,0}^t &=& N\frac{t}{T_{tot}},
\end{eqnarray}
thus maintaining the steady state probability distribution, i.e. zero net in/out-flow. We also note, that in the steady state the number of TCLs in the ON state, $N_1$, and the number of TCLs in the OFF state, $N_0$, can be written as
\begin{eqnarray}\label{N0N1}
N_1 &=& N\frac{T_c}{T_{tot}} \nonumber \\
\& \quad N_0 &=& N\frac{T_h}{T_{tot}}
\end{eqnarray}

We assume that the TCLs were operating in a steady state between deadband limits $\theta_-^0$ and $\theta_+^0$. When the signal to shift the deadband to the right by an amount $\delta$ is received by the TCLs at, say, time instant $t=0$ (Fig. \ref{R1}) the TCLs keep on operating under their current state until they hit one of the ``transition points'', $\theta_-^0$ and $\theta_+=\left(\theta_+^0+\delta\right)$. In the next three sub-sections we will address the different stages through which the shift takes place.

\subsection{Power Profile: $0\leq t \leq \tau_1$}\label{0_tau1}
The steady state aggregate power consumption of the population, at time $t=0$, is
\begin{eqnarray}\label{pss0}
P_{tot}(0) &=& \frac{T_c^0}{T_{tot}}NP \quad (using\;(\ref{N0N1})) \nonumber \\
&=& \frac{T_c^0}{T_{tot}}P_{max}; \quad P_{max} = NP
\end{eqnarray}
where, $T_c^0$ is the original cooling cycle duration, before the shift was initiated. Once the (right) shift process is initiated at $t=0$, the transition points, $\theta_-^0$ and $\theta_+=\left(\theta_+^0+\delta\right)$, become operative. Hence TCLs that were in the OFF state do not switch to ON state until a time $\tau$ which is same as the time taken by a TCL to heat up from $\theta_+^0$ to $\theta_+$, given by
\begin{eqnarray}\label{tau1}
\tau_1 = CR\;ln\left(\frac{\theta_{amb}-\theta_+^0}{\theta_{amb}-\theta_+}\right)
\end{eqnarray}
The TCLs that were in the ON state at $t=0$, however, keep on switching to OFF state as they hit the transition point $\theta_-^0$. Hence, assuming that $\delta$ is small enough to ensure $\tau_1 \leq T_c^0$, there will be a {\it net out-flow} of TCLs from ON to OFF during the time $t\in(0,\tau_1]$. The number of TCLs going from ON to OFF between time $t=0$ and $t=t$ is $\Delta N_{1,0}^t = tN/T_{tot}$ (from (\ref{flow_10})). Assuming the total cycle duration $T_{tot}$ to remain {\it almost} unchanged throughout the shift process (from (\ref{eta_T})), and using (\ref{pss0}) and (\ref{flow_10}), we have
\begin{eqnarray}\label{ptot_0_tau1}
P_{tot}(t) &=& \frac{\left(T_c^0-t\right)}{T_{tot}}P_{max}, \quad \forall t\in(0,\tau_1]
\end{eqnarray}

\subsection{Power Profile: $\tau_1< t \leq \tau_2$}\label{tau1_tau2}
After $t=\tau_1$, TCLs come into the ON state at a rate $N/T_{tot}$ (from (\ref{flow_01})) and go out of the ON state at the same rate $N/T_{tot}$ (from (\ref{flow_10})), ensuring that the net out-flow (or in-flow) in the ON state remains zero (Fig. \ref{R3}). This continues to happen until all the TCLs that were originally ON (at $t=0$) make the switch to OFF state, at time $t=\tau_2$. Since $T_c^0$ is the original cooling cycle duration, clearly
\begin{eqnarray}\label{ptot_tau1_tau2}
\tau_2 &=& T_c^0 \nonumber \\
\& \quad P_{tot}(t) &=& P_{tot}(\tau_1), \quad \forall t\in(\tau_1,\tau_2] \nonumber \\
&=& \frac{\left(T_c^0-\tau_1 \right)}{T_{tot}}P_{max}, \quad \forall t\in(\tau_1,\tau_2]
\end{eqnarray}

\subsection{Power Profile: $\tau_2< t \leq \tau_3$}\label{tau2_tau3}
At $t=\tau_2$ all the originally ON TCLs make the switch to OFF state. Beyond $t=\tau_2$ (Fig. \ref{R4}) there is in-flow of TCLs to ON state at a rate $N/T_{tot}$ but no out-flow, thereby resulting in an increase in the total power consumption. This continues till time $t=\tau_3$ when the current ($t=\tau_2$) ON state loads start hitting the new lower deadband limit $\theta_-=\theta_-^0+\delta$. Since starting at $t=\tau_1$, it takes time $T_c$ (new cooling cycle duration) for the TCLs to start switching from ON to OFF, again, we have
\begin{eqnarray}\label{ptot_tau2_tau3}
\tau_3 &=& \tau_1 + T_c \nonumber \\
\& \quad P_{tot}(t) &=& P_{tot}(\tau_2) + \frac{\left(t-\tau_2\right)}{T_{tot}}P_{max} \quad \forall t\in(\tau_2,\tau_3] \nonumber \\
&=& \frac{\left(T_c^0-\tau_1+t-\tau_2\right)}{T_{tot}}P_{max} \quad \forall t\in(\tau_2,\tau_3] \nonumber \\
\end{eqnarray}

\subsection{Power Profile: $t>\tau_3$}\label{tau3}
Fig.~\ref{R5} shows the situation at a time $t>\tau_3$, where the TCLs are operating according to their new steady state probability distributions, and the aggregate power consumption is
\begin{eqnarray}\label{pss}
P_{tot}(t) &=& P_{tot}(\tau_3) \quad \forall t>\tau_3 \nonumber \\
&=& \frac{T_c}{T_{tot}}P_{max} \quad \forall t>\tau_3 \quad (using\;(\ref{ptot_tau1_tau2})\;\&\;(\ref{ptot_tau2_tau3})) \nonumber \\
\end{eqnarray}

This completes the derivation of $P_{tot}(t)$ for $\delta >0$ and $\tau_1\leq T_c^0$. 

\subsection{Other Operating Regimes  }
\subsubsection{Large Deadband Shift}
The case $\tau_1>T_c^0$ occurs when $\delta$ is reasonably large, in which situation the total power $P_{tot}(t)$ goes to zero at time $t=T_c^0 < \tau_1$ and thereafter stays at $P_{tot}(t)=0$ until $t=\tau_1+T_h$ at which point the OFF state TCLs start making transition from OFF to ON and so $P_{tot}(t)$ starts to increase linearly before reaching the new steady state at $t=\tau_1+T_h+T_c$ after which total power consumption stays at $P_{tot}(t)=P_{max}T_c/T_{tot}$. 

\subsubsection{$\delta<0$}
The situation when the deadband is shifted to left ($\delta<0$) can be analyzed quite similarly. Here we only present the final form that $P_{tot}(t)$ is going to have
\begin{eqnarray}\label{shift_left}
P_{tot}(t) &=& \left\{ \begin{array}{cl} P_{max}{T_c^0}/{T_{tot}}, & \forall t\leq0  \\ P_{max}\left(T_c^0+t\right)/{T_{tot}} & \forall t\in(0,\tau_1 ']   \\ P_{max}\left(T_c^0+\tau_1 '\right)/{T_{tot}} & \forall t\in(\tau_1 ', \tau_2 ']  \\ P_{max}\left(T_c^0+\tau_1 ' - t + \tau_2 '\right)/{T_{tot}} & \forall t\in(\tau_2 ', \tau_3 ']  \\ P_{max}{T_c}/{T_{tot}} & \forall t>\tau_3' \end{array} \right. \nonumber \\
\end{eqnarray}
where, $\tau_1 '=CR\;\ln\left(\frac{PR+\theta_-^0-\theta_{amb}}{PR+\theta_- -\theta_{amb}}\right), \tau_2 ' = T_h^0, \tau_3 ' = \tau_1 '+ T_h$ and $T_{tot}=T_c^0+T_h^0\approx T_c+T_h$.

\subsection{Total Energy Consumed in the Process}\label{energy_consumption}
At $t<0$, the population of TCLs consumes energy at a rate $P_{tot}(0)=P_{max}T_c^0/T_{tot}$, in (\ref{pss0}), and after $t>\tau_3$ at a rate $P_{tot}(\tau_3)=P_{max}T_c/T_{tot}$. During the time interval $[0,\tau_3]$ the total energy consumed can be expressed as 
\begin{eqnarray}
E_{0,\tau_3} &=& \frac{1}{2}\tau_1 \left(P_{tot}(0)+P_{tot}(\tau_1)\right) + \left(\tau_2-\tau_1\right) P_{tot}(\tau_1) \nonumber \\ 
						 && + \frac{1}{2} \left(\tau_3-\tau_2\right) \left(P_{tot}(\tau_2)+P_{tot}(\tau_3)\right) \nonumber \\
&=& \frac{\left(\left(T_c^0\right)^2+T_c^2\right)}{2T_c^0}P_{tot}(0) \nonumber \\
&=& \frac{\left(\left(T_c^0\right)^2+T_c^2\right)}{2T_{tot}}P_{max} \nonumber \\
\end{eqnarray}

\section{NUMERICAL RESULTS}\label{numerical}
\begin{figure*}[thpb]
\centering
\subfigure[]{
\includegraphics[width=2in]{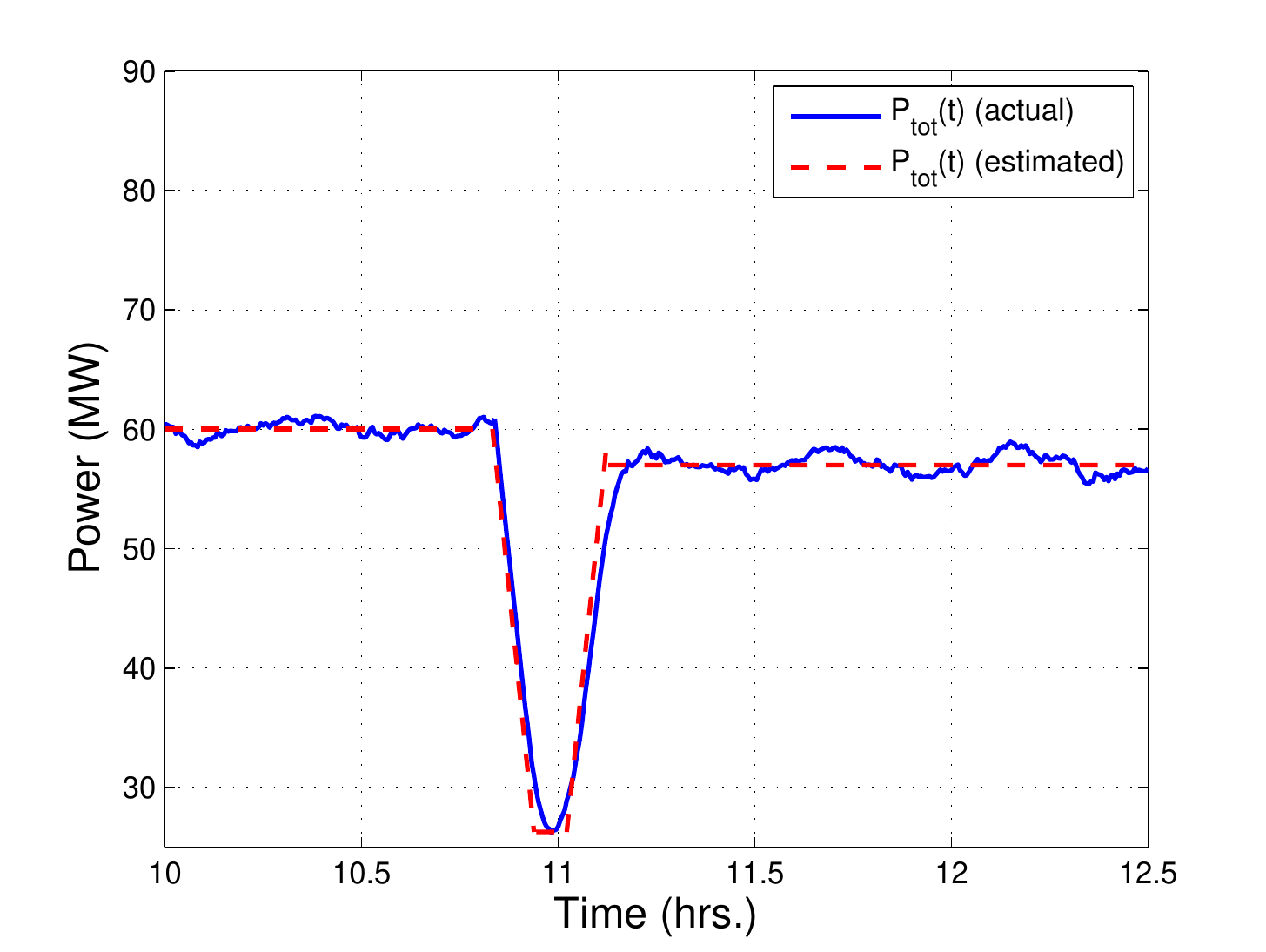}\label{safe_R}
}\quad
\subfigure[]{
\includegraphics[width=2in]{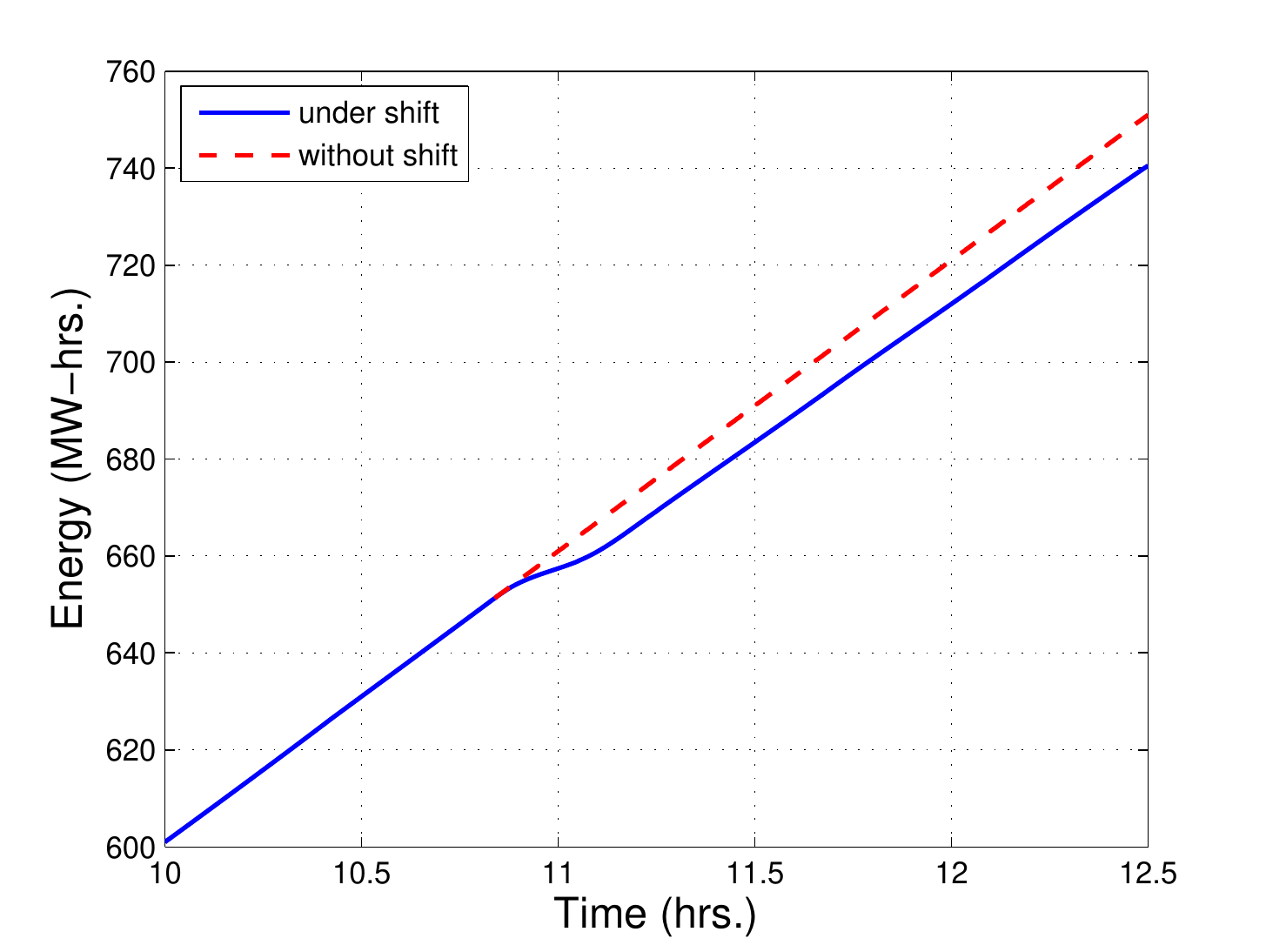}\label{safe_R_E}
}\quad
\subfigure[]{
\includegraphics[width=2in]{input_u0d60.pdf}\label{input_R_safe}
}
\caption[Optional caption for list of figures]{Power and energy response under a ``safe'' shift to higher setpoint temperature.}
\label{safe_increase}
\end{figure*}
\begin{figure*}[thpb]
\centering
\subfigure[]{
\includegraphics[width=2in]{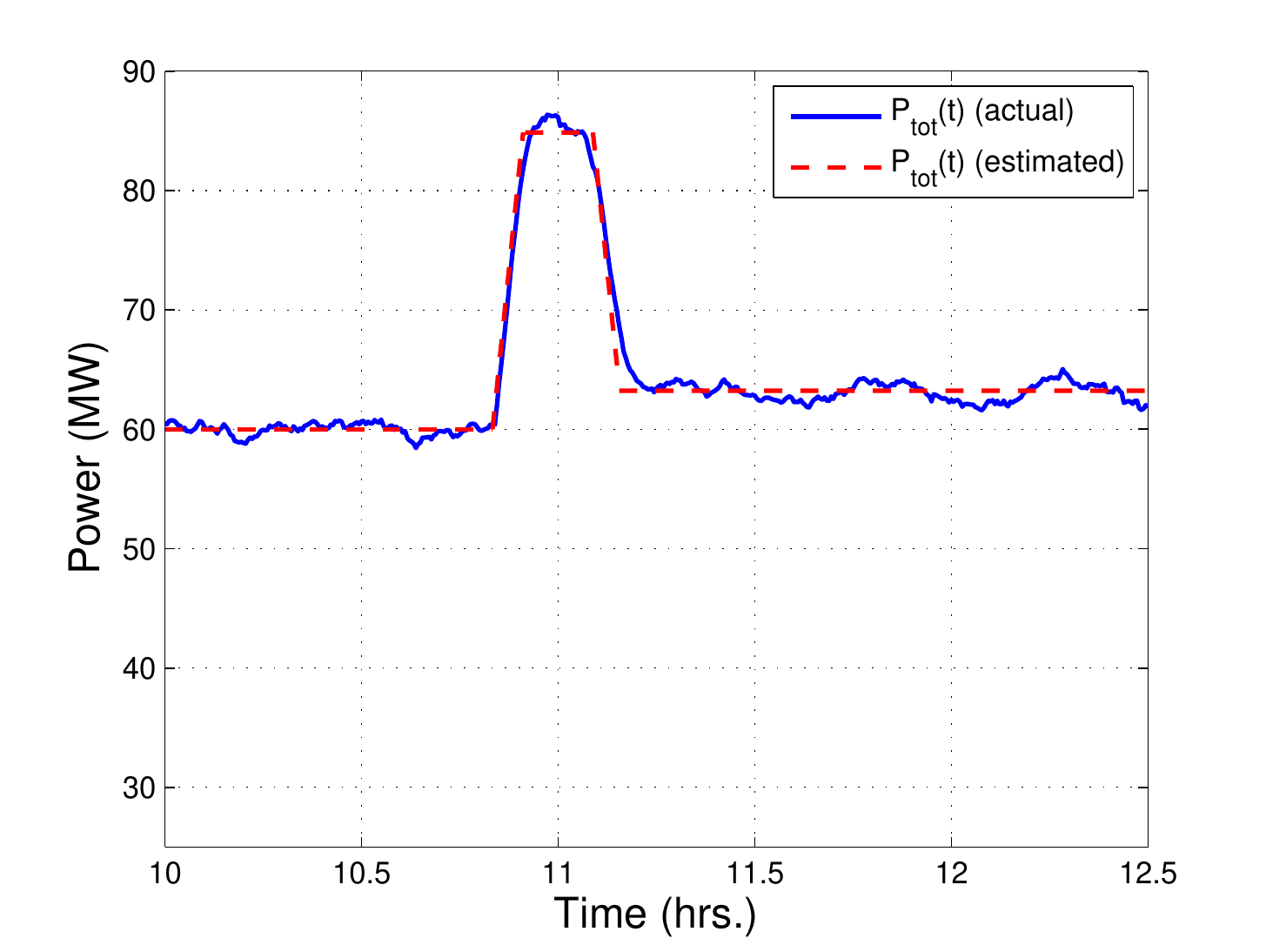}\label{safe_L}
}\quad
\subfigure[]{
\includegraphics[width=2in]{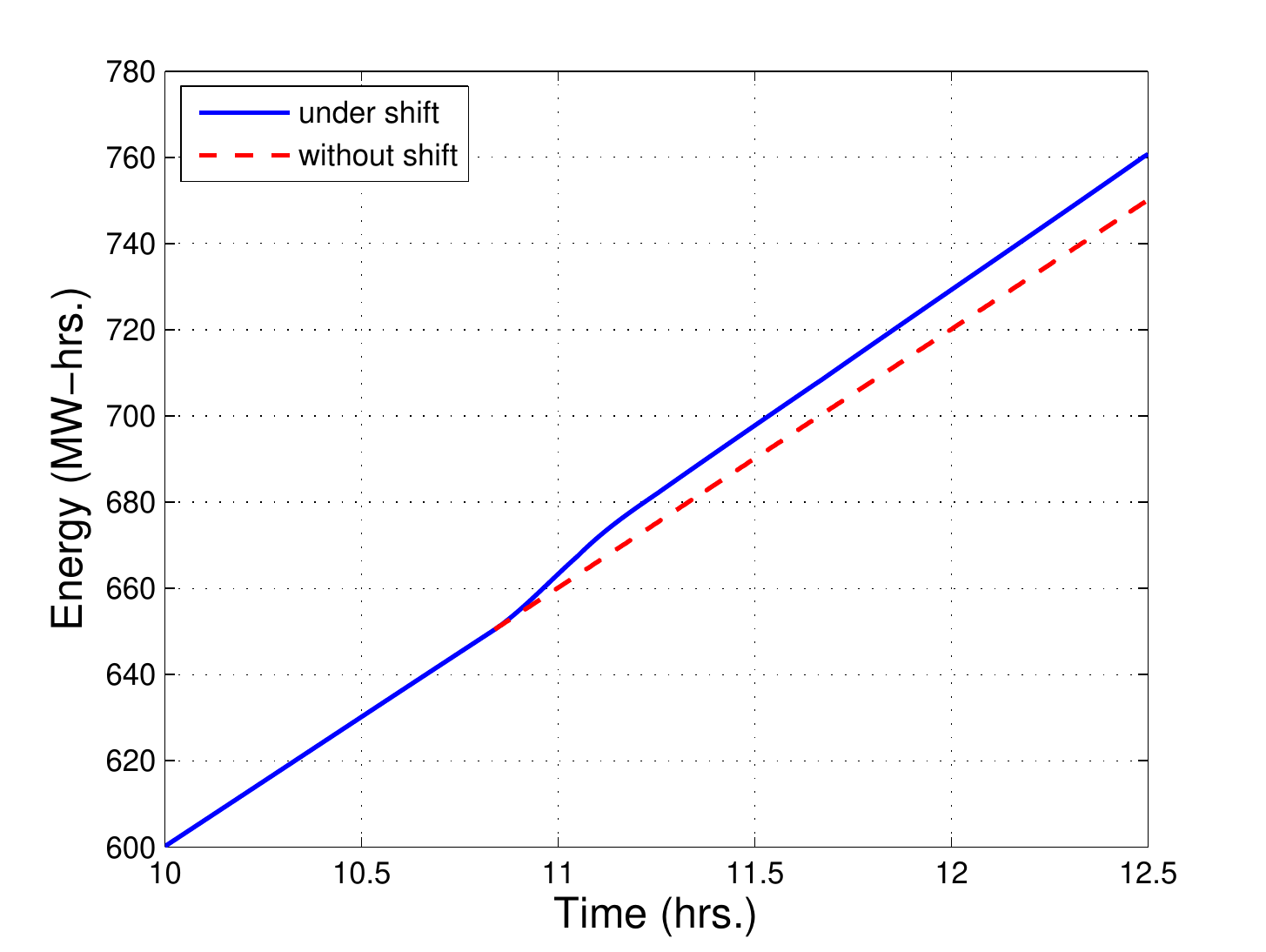}\label{safe_L_E}
}\quad
\subfigure[]{
\includegraphics[width=2in]{input_u0d-60.pdf}\label{input_L_safe}
}
\caption[Optional caption for list of figures]{Power and energy response under a ``safe'' shift to lower setpoint temperature.}
\label{safe_decrease}
\end{figure*}
\begin{figure}[thpb]
\centering
\subfigure[]{
\includegraphics[width=2.5in]{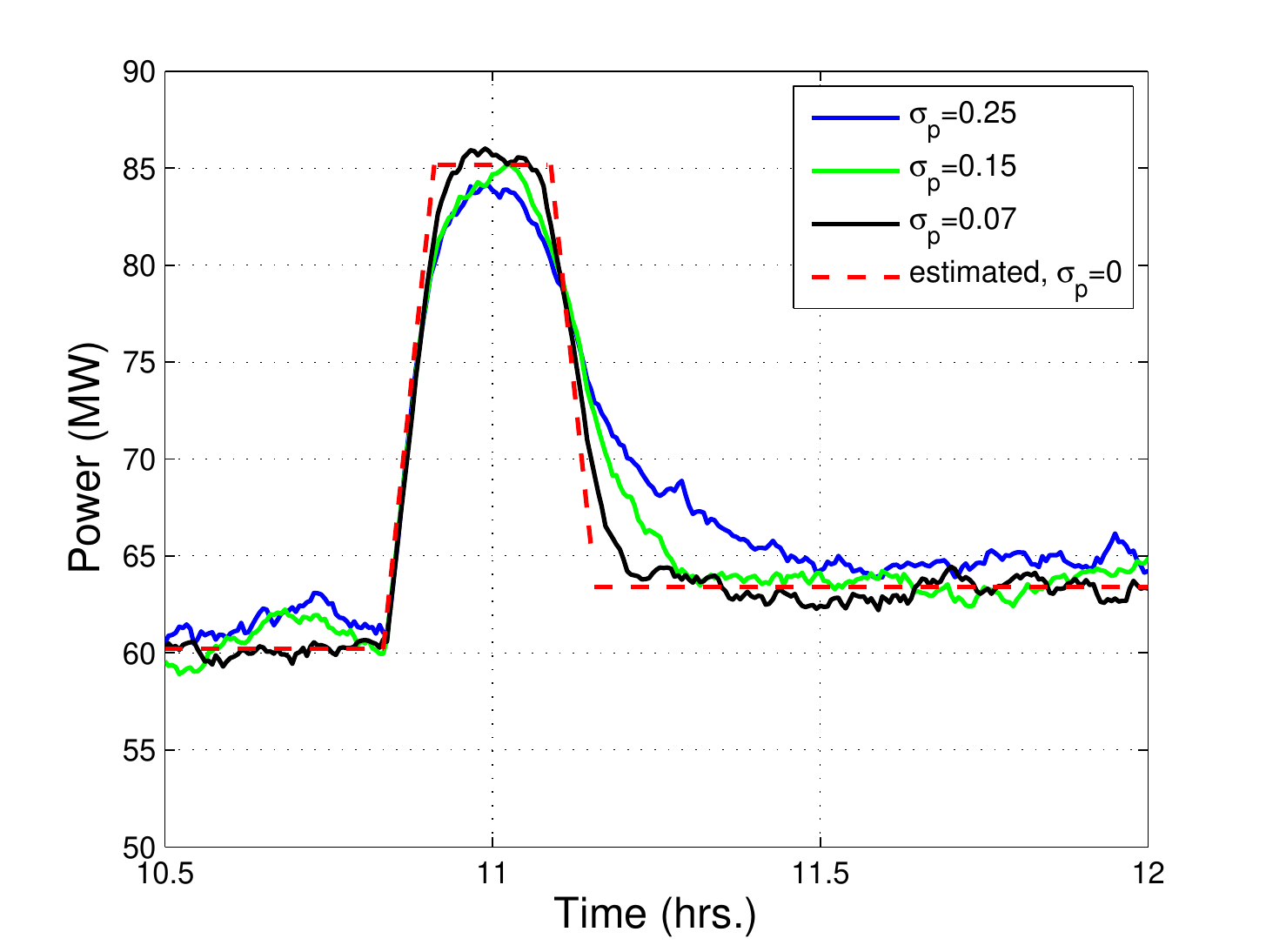}\label{safe_sigma_p}
}
\caption[Optional caption for list of figures]{Performance of ``safe protocol'' to input in Fig.~\ref{input_L_safe}, under varying heterogeneity.}
\label{safe_hetero}
\end{figure}

Figs.~\ref{safe_increase}-\ref{safe_decrease} illustrate the effect of application of the light version of the safe protocol on an ensemble of TCLs (parameters given in Section~\ref{problem_statement}) under step changes in setpoint temperature. The large oscillations, visible otherwise in Figs.~\ref{normal_increase}-\ref{normal_decrease}, disappear and the population attains the new steady state within the time period $T_{tot}$ of the TCL's cycle. It also results in a steady change in energy consumption within a time period duration. Fig.~\ref{safe_hetero} shows how this safe protocol fares under increasing heterogeneity measured by the standard deviation of the lognormal distribution of paramaters ($C,R$ and $P$), which is a factor, $\sigma_p$, of corresponding mean. As $\sigma_p$ increases, the actual response deviates further away from theoretically estimated response. But most importantly, the large fluctuations do not appear and the power response settles to the new steady state within a time period duration.

\section{CONCLUSION}\label{conclusion}
We proposed a method to implement simultaneous shifts of the  deadband positions in a large heterogeneous population of TCLs without inducing parasitic power oscillations in the power grid. 
Our method is helpful in getting rid of the unwanted oscillations at the onset of a deadband shift and could be a tool in demand side energy management. Using the knowledge of the energy consumed in a deadband shift under the safe protocol a population of TCLs could potentially be used to compensate for the over/under-generation of renewable energy within a relatively short duration, about the time period of the operation of a TCL. Potentially, nonperturbative control of TCLs can be equivalent to the additional possibility of manipulating the power of up to $100~MW$ in a city with a million houses. While this method is concerned with load-side energy management, suitable feedback control may be designed to utilize the absence of parasitic oscillations and track the fluctuations in generation.

\vspace{0.1in}

\addtolength{\textheight}{-3cm}   


\section{ACKNOWLEDGMENTS}
We would like to thank Prof. Ian Hiskens of University of Michigan, Ann Arbor, USA, Dr. Scott Backhaus and Dr. Michael Chertkov of Los Alamos National Laboratory, USA for their useful insights at various stages of this work. This work was supported by DOE under Contract No. $DE$-$AC52$-$06NA25396$.

\end{document}